\documentclass[12pt]{article}
\pdfoutput=1
\usepackage{float}
\usepackage[final]{graphicx}
\usepackage{amssymb,amsmath}
\usepackage{commath}
\usepackage{epsfig}
\usepackage{epstopdf}
\usepackage{latexsym}
\usepackage{graphicx}
\usepackage{subfigure}
\usepackage{booktabs}
\usepackage{tkz-euclide}

\usepackage{makeidx}
\usepackage{cite}
\usepackage{bm}
\usepackage{geometry}
\usepackage{braket}
\usepackage[margin=20pt,small]{caption}

\usepackage[toc]{appendix}

\usepackage{tikz}
\usetikzlibrary{calc,decorations.markings,arrows.meta,shapes.misc,decorations.pathmorphing,calc,bending}

\usetikzlibrary{arrows.meta,shapes.misc,decorations.pathmorphing,calc,bending}

\tikzset{
  branch point/.style={cross out,draw=black,fill=none,minimum size=2*(#1-\pgflinewidth),inner sep=0pt,outer sep=0pt}, 
  branch point/.default=5
}
\tikzset{
  branch cut/.style={
    decorate,decoration=snake,
    to path={
      (\tikztostart) -- (\tikztotarget) \tikztonodes
    },
    }
  }

\usepackage{color}
\usepackage{datetime}
  
\ifpdf
\fi

\def\O{{\cal O}}

\definecolor{cardinal}{rgb}{0.6,0,0}
\definecolor{darkgreen}{rgb}{0,0.5,0}
\definecolor{golden}{rgb}{0.92, 0.7, 0}
\definecolor{midnight}{rgb}{0, 0, 0.5}
\definecolor{darkblue}{rgb}{0.2, 0, 0.8}

\newcommand{\be}{\begin{equation}}
\newcommand{\ee}{\end{equation}}
\newcommand{\bea}{\begin{eqnarray}}
\newcommand{\eea}{\end{eqnarray}}

\begin{document}

\begin{titlepage}

\bigskip
\bigskip
\bigskip
\centerline{\Large \bf Critical Quenches, OTOCs and Early-Time Chaos}
\bigskip
\centerline{\bf Suchetan Das$^{1}$, Bobby Ezhuthachan$^2$, Arnab Kundu$^{3}$, Somnath Porey$^2$, Baishali Roy$^2$}
\bigskip
\bigskip
\centerline{$^1$Department of Physics,}
\centerline{Indian Institute of Technology Kanpur,}
\centerline{Kanpur 208016, India.} 
\bigskip
\bigskip
\centerline{$^2$Ramakrishna Mission Vivekananda Educational and Research Institute,}
\centerline{Belur Math,}
\centerline{Howrah-711202, West Bengal, India.} 
\bigskip
\bigskip
\centerline{$^3$ Theory Division, Saha Institute of Nuclear Physics,}
\centerline{Homi Bhaba National Institute (HBNI),}
\centerline{1/AF, Bidhannagar, Kolkata 700064, India.}
\bigskip
\bigskip
\centerline{suchetan[at]iitk.ac.in,  bobby.ezhuthachan[at]rkmvu.ac.in, arnab.kundu@saha.ac.in}
\centerline{somnathhimu00[at]gm.rkmvu.ac.in, baishali.roy025[at]gm.rkmvu.ac.in  }
\bigskip
\bigskip
\bigskip

\begin{abstract}

\noindent In this article, we explore dynamical aspects of Out-of-Time-Order correlators (OTOCs) for critical quenches, in which an initial non-trivial state evolves with a CFT-Hamiltonian. At sufficiently large time, global critical quenches exhibit a universal thermal-behaviour in terms of low-point correlators. We demonstrate that, under such a quench, OTOCs demarcate chaotic CFTs from integrable CFTs by exhibiting a characteristic exponential Lyapunov growth for the former. Upon perturbatively introducing inhomogeneity to the global quench, we further argue and demonstrate with an example that, such a perturbation parameter can induce a parametrically large scrambling time, even for a CFT with an order one central charge. This feature may be relevant in designing measurement protocols for non-trivial OTOCs, in general. Both our global and inhomogeneous quench results bode well for an upper bound on the corresponding Lyapunov exponent, that may hold outside thermal equilibrium.

\end{abstract}

\newpage


\end{titlepage}
\tableofcontents


\rule{\textwidth}{.5pt}\\

\section{Introduction}

Ergodicity is one of the main cornerstones of modern understanding of statistical mechanics and the dynamics of thermalization. While classical notion of ergodicity can be cleanly defined in terms of the phase-space of a system, in the quantum regime, this becomes subtle. The conventional measure of ergodicity or chaos in a quantum system is in terms of the energy-level statistics of the corresponding spectrum. In particular, level-repulsion is a tell tale sign of a quantum chaotic system\cite{Stockmann}.

On the other hand, given a quantum dynamics, various time-scales demarcate qualitatively distinct physics. The imprint of a level-repulsion appears on the real-time dynamics at the largest possible time-scale, the {\it Heisenberg time} $t_{\rm H} \sim \Delta^{-1}$, where $\Delta$ is the mean level-spacing and provides the smallest scale in the system. See {\it e.g.}~\cite{Altland:2020ccq} for a recent review on various time-scales in quantum dynamics. For a wide class of systems, quantum chaotic physics can appear at a much shorter time-scale, known as the scrambling time $t_*\sim \log N$, where $N$ is a parametrically large number, {\it e.g.}~the number of degrees of freedom in the system. While the former can be associated with a {\it late-time chaos}, the latter is understood as an {\it early-time chaos}. In this article, we will focus on the latter.

Early-time chaos can be diagnosed in a special class of correlation functions, the Out-of-Time-Order Correlators, OTOCs in short. Typically, an exponential piece in such a $4$-point OTOC of a pair of operators translate into exponential growth in the expectation value of the commutator-squared of the corresponding pair. In the semi-classical limit, $\hbar \to 0$, this intuitively parallels the exponential sensitivity of classical trajectories in the corresponding classical system. For thermal states with a temperature $T$, this behaviour is visible above the dissipation time-scale $t_{\rm d} \sim T$.

This class of correlators have recently found numerous new applications, ranging from the physics of disordered systems to the physics of black holes, see {\it e.g.}~\cite{Swingle2018} for a review. Moreover, remarkable recent progress in quantum control of atoms and ions have picked an active interest in proposing measurement protocols for OTOCs ({\it e.g.}~\cite{Zhu:2016uws, Swingle2016}) and early measurements on trapped-ion systems ({\it e.g.}~\cite{Gartner2017}), which are closely related to the so-called Loschmidt echoes in spin systems and involve a quantum time-reversed evolution of the same.\footnote{See also \cite{Yoshida:2018vly, Landsman:2018jpm} for recent suggested protocols for measuring OTOC in the lab.}

While exciting progress is taking place in the experimental front, rather limited theoretical control is available on OTOCs in general, specially outside thermal states in systems with a large number of degrees of freedom. The best understood examples are in two-dimensional conformal field theories (CFTs) with a large central charge\cite{Roberts:2014ifa}, thermal states in CFTs with a Holographic dual in arbitrary dimensions\cite{Maldacena:2015waa}, and some examples on thermal states in weakly-coupled large-$N$ gauge theories\cite{Stanford:2015owe, Steinberg:2019uqb}. All these examples warrant a thermal state and a large-$N$ system. While thermal states are common,\footnote{At least in an approximate sense for a typical state in the Hilbert space. This is quantitatively precise for an energy eigenstate satisfying the Eigenstate Thermalization Hypothesis.} experimentally, one would likely access a small-$N$ system in which any non-trivial OTOC dynamics will only be transient.

In this article, we generalize both aspects. First, we move away from a typical state and consider a case when the state is not in an {\it ab initio} equilibrium. Secondly, we introduce a small parameter in the system,  which makes it possible to have non transient chaotic behaviour in some finite $N$ examples, albeit in a perturbative regime.\footnote{Note that, the semi-classical notion of chaos comes naturally equipped with a small number, since it generalizes the notion of sensitivity of classical trajectories under a {\it small} change in the initial condition. This can be set by $(1/N)$ of a large-$N$ system, $\hbar$ in an appropriate unit. In this article, we will see that coupling the system with an external non-dynamical field can also achieve this.}

We discuss these issues within the framework of the quantum quench protocol\cite{bm1, ir-00, sps-04} , which provides a simple setup to study non equilibrium dynamics and thermalization in isolated quantum systems. The quantum quench dynamics, as the name suggests, refers to a process where the  Hamiltonian of a system is changed over a very short time scale, for instance by a sudden tuning of some of its parameters\footnote{In this paper we assume the dynamics to a global quench, where global refers to the fact that the initial state and eigenstate of the quenched hamiltonian differ globally, as opposed to local quenches where the difference is local\cite{Calabrese:2007mtj, StDu:1105}.}. One then studies time evolution of correlation functions of local operators and entanglement structure of the system following the quench. Interest in understanding the dynamics of quantum systems following a quench, has received an impetus since the seminal experiments involving ultracold atoms, where such a protocol was realized between superfluid and mott-insulating states\cite{uc}.

We will focus on  systems, where the dynamics after the quench is a CFT.  In\cite{Calabrese:2006rx}, Cardy and Calabrese pioneered the study of the  quench dynamics in these systems, and in particular computed the late time behaviour of one point and two point functions of primary operators. The initial state of the system was chosen to be a very special state --- the  so called Cardy-Calebrese state(CC), which is essentially a conformal boundary state of the CFT, suitably regulated to have a finite norm. There are two fold advantages in choosing the CC state. Firstly, the computation of correlation function in the CC state is equivalent  to a BCFT computation which makes the computation analytically tractable. Secondly, as argued in \cite{Calabrese:2006rx}, the CC state may be physically thought of as approximating the  ground state of a Hamiltonian which is an irrelevant deformation of the CFT. Moreover, the regulator parameter $\tau_0$ maybe interpreted as the correlation length of this ground state.  By computing one point and two point 
functions of primary operators  in this state, the authors show that at late times, the results are identical to the ones in a thermal state, with the effective temperature being set by $\tau_0$. In effect, the CC state self-thermalizes at late times. This result is universal for all CFT's, so it holds for integrable as well as chaotic CFTs. This setup can be further generalized, for instance  by introducing a spatial inhomogeneity in the original hamiltonian\cite{Sotiriadis:2008ila}. This could be done by adding impurities into the system which breaks the translational invariance of the system. This would effectively mean the correlation length of the ground state is now position dependent. In the  critical quench scenario, this can be implemented by making $\tau_0(x)$.

  In this paper, we study the late time behaviour of the OTOC of primary operators in the CC state as well as in a perturbed CC state in different examples of CFTs, including superintegrable, weakly integrable and chaotic CFTs. As examples of each we study the OTOC in a minimal model CFT, orbifold CFT and a large-$c$ CFT  respectively. 
  
  Firstly, we show that under a critical quench dynamics in $(1+1)$-dimensions, in a large-$c$ CFT$_2$, the corresponding OTOC still exhibits a chaotic behaviour similar to that of a thermal state --- with an effective temperature set by $\tau_0$. Note that this result is non-trivial since it captures physics away from standard CFT-universality in two and three point functions. On the other hand, the effective thermal intuition ties naturally with the universal quench dynamics observed in the lower point correlations of  \cite{Calabrese:2016xau}, and selects out a class of 2D CFTs where this notion is much stronger.  It is also interesting to place this result in light of a general effective thermal physics in 2D CFT with heavy-states, see {\it e.g.}~\cite{Anous:2019yku, Anous:2020vtw, Kundu:2021bub}

Secondly, we also demonstrate how an inhomogeneous critical quench, with a small inhomogeneity parameter that serves the role of an external field, can provide a scrambling window within which an exponential growth in OTOC can be distilled. Note that, it is not enough to have a small parameter in the system to warrant the hierarchy $t_* \gg t_{\rm d}$. For example, a weakly coupled field theory cannot do this, despite having a small coupling constant. On the contrary, to extract the Lyapunov growth of OTOCs, in the weakly coupled regime, one needs to carry out a perturbation series in high-orders of the coupling constant and subsequently resum them. Thus, the final result becomes, in a certain sense, non-perturbative in the coupling constant. See {\it e.g.}~\cite{Stanford:2015owe, Steinberg:2019uqb} where such weakly coupled results are obtained, necessarily with a large-$N$ system.

Our results show that the small inhomogeneity parameter $a$, which is supplied from outside, can indeed provide us with the hierarchy. In fact, we show with an example that an integrable 2D CFT with $c=2$,  subject to an  inhomogeneous quench, indeed develops a chaotic exponential growth supported by this small parameter. Subsequently, the scrambling time can be separated from the scale $\tau_0$ by a factor of $(-\log a)$.

This article is divided into the following parts:  In section \ref{sec2} we give a brief review of the quench set up in CFT$_2$ \cite{Calabrese:2016xau}. Section \ref{sec3} and \ref{sec4} contains the bulk of our analysis and results. We consider two complimentary limits in the paper. In section \ref{sec3} we  study the limit $\tau_0\rightarrow 0$. After setting up the basic framework in section \ref{sec3.1}, we study a special 3-pt OTOC in the CC state, for the homogenous quench in a large-$c$ CFT in section \ref{sec3.2}. We explicitly show that at late times, the OTOC shows a maximal lyapunov growth with an effective temperature $\beta=4\tau_0$. We then analyze the same 3-pt OTOC for a large-$c$ CFT after introducing a small inhomogeneity in the quench setup in section \ref{sec3.3}.  We show, that at leading order in the perturbation, the lyapunov exponent remains the same, though there is a change in the butterfly velocity. In section \ref{sec4}, we focus on the opposite large $\tau_0$ limit and study a 4-pt OTOC for three examples- the large-$c$ CFT (section \ref{sec4.1}), minimal model CFT (section \ref{sec4.2}) and an orbifold CFT (section \ref{sec4.3}). For the case of the large-$c$ CFT, we show that in the presence of the inhomogenous perturbation, the effective Lyapunov exponent changes due to the perturbation.  In the case of the minimal model CFT, as expected, the perturbation does not change the nature of the OTOC, and thus does not induce chaotic behaviour. On the other hand, for an orbifold CFT, we see a chaotic growth in the OTOC in the presence of inhomogeneity, within a perturbative regime. Moerover, the small inhomogeneity parameter induces a large hierarchy between $\tau_0$, and the scrambling time. We end with a summary of the key conclusions of our paper and a discussion of some future directions in section \ref{sec5}.

\numberwithin{equation}{section}
\section{Correlation functions: General set up}\label{sec2}

Vacuum correlation functions in a Lorentzian QFT are typically defined by an analytic continuation in the time coordinate via the $i\epsilon$ prescription. ($t\rightarrow \tau =t-i\epsilon$).
\begin{equation}
\langle \Omega |\phi_1({\bf x}_1, t_1)\phi_2({\bf x}_2, t_2)\cdots \phi_n({\bf x}_n,t_n)|\Omega\rangle   \equiv \langle \Omega |\phi_1({\bf x}_1)e^{-iH\tau_{12}}\phi_2({\bf x}_2)e^{-iH\tau_{23}}\cdots e^{-iH\tau_{(n-1)n}}\phi_n({\bf x}_n)|\Omega \rangle |_{\epsilon_{i} \rightarrow 0} 
\end{equation}
Where $\tau_{ij}= \tau_i -\tau_j$ and the Hamiltonian $H$ is bounded from below (assumed here for simplicity to be a positive semi-definite  operator). The $e^{-H\epsilon_{ij}}$ factor provides a UV regulator as long as $\epsilon_{ij}> 0$. Thus the RHS is well-defined provided ($\epsilon_1 >\epsilon_2>\cdots >\epsilon_n$). This procedure however fails if we compute correlation functions in an arbitrary state (say $|\Psi_0\rangle$), which is not an eigenstate of the Hamiltonian, since  by the same argument as given above,  
\begin{equation}
\langle \Psi_0 |e^{iH\tau_1}\phi_1({\bf x}_1)e^{-iH\tau_{12}}\phi_2({\bf x}_2)e^{-iH\tau_{23}}\cdots e^{-iH\tau_{(n-1)n}}\phi_n({\bf x}_n)e^{-iH\tau_n}|\Psi_0 \rangle |_{\epsilon_{i} \rightarrow 0} \nonumber
\end{equation} 
is analytic only when ($ 0>\epsilon_1 >\epsilon_2 >\cdots> \epsilon_n > 0 $ ), which is impossible to satisfy. One way to get around this and get a finite analytic region is to introduce a regulator ($\tau_0$) which effectively cuts off the very high energy modes with energy greater than $\frac{1}{\tau_0}$. This may be achieved by substituting the state by the regulated state $|\Psi_0\rangle \rightarrow e^{-\tau_0 H}|\Psi_0\rangle$. Here $\tau_0$ is necessarily non-zero positive real number. In this regulated state, the correlation function maybe defined via an analytic continuation as above, provided we restrict $( \tau_0 > \epsilon_1 >\epsilon_2 > \cdots >\epsilon_n >-\tau_0)$. This essentially means that the Euclidean theory whose analytically continued version is described by the real time correlation function is defined on a strip-ie: All $\tau_i$ lie on the strip ($\tau_i =t_i -i\epsilon_i, \; \infty >t_i>-\infty,\; \;\textrm{and}\;\; |\epsilon_i| < \tau_0 $). Thus in the Euclidean picture, the correlation function computation becomes a computation on a strip of length $2\tau_0$, with the boundary conditions at the ends of the strip being determined by the state $|\Psi_0\rangle$. Note that, this picture is also valid for evaluating Out-of-Time-Order Correlators (OTOCs), for which an ordered $n$-tuple of $\{\epsilon_i\}$ decides the corresponding operator (time)-ordering in the correlator.\footnote{For example, $\tau_0 > \epsilon_1 >\epsilon_2>\epsilon_3 >-\tau_0 $ defines $\langle \Psi_0 | \phi_1(\tau_1, x_1) \phi_2 (\tau_2, x_2) \phi_3(\tau_3,x_3) |\Psi_0 \rangle$, whereas $\tau_0 > \epsilon_2 >\epsilon_1>\epsilon_3 >-\tau_0 $ defines $\langle \Psi_0 | \phi_2(\tau_2, x_2) \phi_1 (\tau_1, x_1) \phi_3(\tau_3,x_3) |\Psi_0 \rangle$. The former one is a Time-Order Correlator (TOC) with $\tau_1>\tau_2>\tau_3$, but the latter one is an OTOC.}

In applications to quantum quench problems, the state $|\Psi_0\rangle$ is taken to be an eigenstate (typically the ground state) of the original Hamiltonian ($H_0$ ) before quenching. In [3], Cardy and Calabrese studied a quantum quench from a massive Hamiltonian $H_0$ to a CFT Hamiltonian $H$. If $H_0$ is close to the CFT Hamiltonian $H$ in the RG sense, ie $H_0$ is obtained from $H$ by a small irrelevant deformation, then [3] argued that correlation functions in such a state over time scales and length scales much larger than the correlation length ($\xi $) of the state, show universal behaviour which may be captured by replacing the state by a regulated conformal boundary state in the fixed point CFT.

Thus in this approximation, we may take the state $|\Psi_0\rangle$ to be the conformal boundary state in the BCFT on a strip. The regulator $\tau_0$ now has a physical interpretation as the  the correlation length of the state in consideration ie: ($\tau_0\sim \xi $). Thus the Euclidean correlation function calculation reduces to a BCFT computation on a strip. 

This set-up may be generalized to incorporate the case when the state in consideration has a spatially inhomogeneous profile (ie $\xi= \xi(x)$). At least when the inhomogeneity is small, this can be modeled by replacing the constant width Euclidean strip $2\tau_0$ by a strip with a position dependent width $2\tau_0(x)$.

In CFT, the 'variable width strip' geometry can be mapped, via a suitably chosen conformal transformation, to a 'constant width strip' geometry, thus reducing the problem to the homogenous quench. Following the notation of the original work, we refer to the `variable width strip' geometry as (VWS) and the `constant width strip' as (CWS). Let the coordinate on VWS be $\omega = x+i\tau$ where at any value of $x$, $\tau$ ranges from $-\tau_0(x)$ to $\tau_0(x)$ and that of CWS be $\zeta = \tilde{x}+i\tilde{\tau}$ where $-\tau_0<\tilde{\tau}<\tau_0$. Then the conformal map from VWS to CWS, $g:\omega \rightarrow \zeta = g(\omega)$, must satisfy the boundary condition
\begin{align}\label{boundary cond g}
\textrm{Im} g(x\pm i\tau_0(x)) = \pm \tau_0
\end{align}

For the simplified case where the inhomogeneity is taken to be a small fluctuation over $\tau_0$, $ie:$ $\tau_0(x) = \tau_0 +h(x)$ with $h(x) \ll \tau_0$, an explicit solution for the map $g(\omega)$, or more precisely its inverse has been obtained in \cite{Sotiriadis:2008ila}. Parameterizing the infinitesimal map as $g^{-1}(\zeta)\equiv \zeta +f(\zeta)$, one finds that the inverse map must satisfy the following boundary condition:\footnote{ This condition already makes use of the fact that $f$ is small, otherwise the argument inside $f$ should have $\tilde{x}$, instead of $x$.}
\begin{align}
{\rm Im} f(x \pm i\tau_0) = \pm h(x) \ .
\end{align}
If one further assumes that the VWS geometry has a reflection symmetry about $\tau=0$\footnote{Which further implies $\overline{g(\omega)}=g(\bar{\omega})$ and hence $\overline{f(\omega)}=f(\bar{\omega})$. See \cite{Sotiriadis:2008ila} for more details.}, then $f(\zeta)$ can be determined, upto an irrelevant real constant, in terms of the function $h(x)$ as: 
\begin{align}\label{f(x)}
f(\zeta) =\frac{1}{\tau_{0}} \int^{\infty}_{-\infty}ds \frac{1}{e^{-\frac{\pi (\zeta-s)}{\tau_0}}+1} h(s) \ .
\end{align}
Here $h(x)$ has been normalized to satisfy $h(-\infty)=0$. An interesting limit, which is useful in this quench scenario, is taking $\tau_0 \rightarrow 0$. In this limit, \ref{f(x)} reduces to the following
\begin{align}\label{asymptotic}
\lim_{\tau_0 \rightarrow 0}f(x) = \frac{1}{\tau_0}\int^{\infty}_{-\infty}ds \theta(\zeta-s) h(s) = \frac{1}{\tau_0} \int^{x}_{-\infty}ds h(s) \ . 
\end{align}
Correlation functions of primary operators in the VWS geometry is related to that of the CWS geometry in the standard way.
\begin{equation}\label{V2C}
\langle \phi'_1(\omega_1, \bar{\omega}_1)\cdots \phi'_n(\omega_n, \bar{\omega}_n) \rangle |_{\rm VWS} =  \prod^{n}_{i=1}\left(\frac{\partial \zeta_i}{\partial \omega_i}\right)^{h_i}\left(\frac{\partial \bar{\zeta}_i}{\partial\bar{\omega}_i}\right)^{\bar{h}_i}\langle \phi_1(\zeta_1, \bar{\zeta}_1)\cdots \phi_n(\zeta_n, \bar{\zeta}_n)\rangle|_{\rm CWS}  \ . 
\end{equation}
 In the next section, we will study the OTOC in the inhomogeneous quench set up in two extreme limits $\tau_0\rightarrow 0$ and the opposite large $\tau_0$ limit. 
 
 \numberwithin{equation}{section}
\section{OTOC in the inhomogeneous quench system}\label{sec3}

\subsection{The Basic Framework}\label{sec3.1}

\begin{figure}[h]
\centering
\includegraphics[scale=0.25]{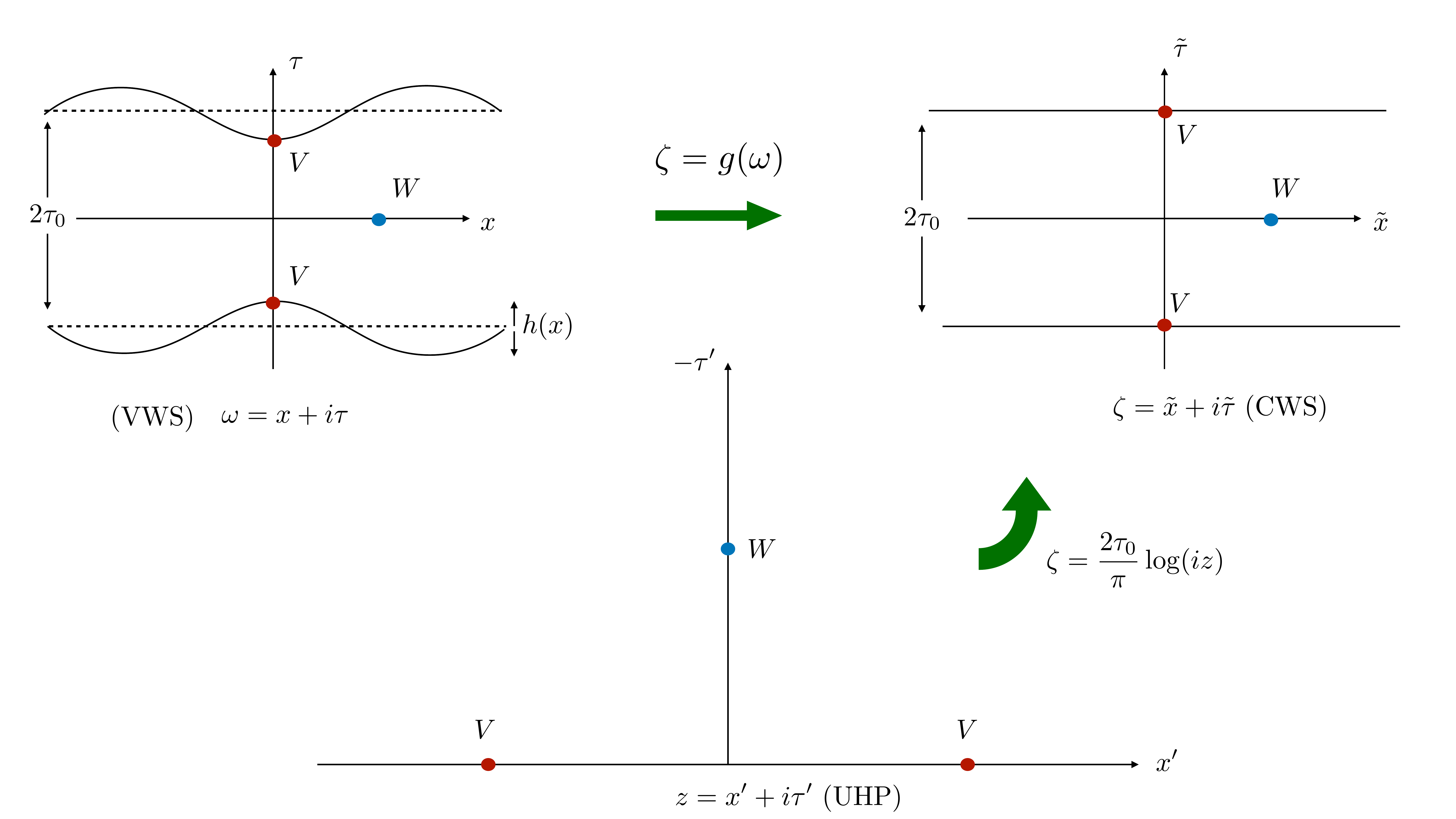}
\caption{The conformal maps from VWS to CWS to UHP. The explicit operator configuration, corresponding to the $b$-OTOC, is given in the UHP where two operators lie on the boundary and one operator is located inside the bulk. }
\label{map plot}
\end{figure}

The first case we study is a bulk-boundary three point OTOC function on the VWS. It was demonstrated in \cite{Das:2019tga} that this class of $3$-point correlator capture OTOC-dynamics and from hereon we refer to them as $b$-OTOC. The scalar operator $W$ of dimension $h_{W}(=\bar{h}_{W})$ is placed at $(x,0)$ and two boundary scalar operators $V$ of dimension $h_{V}$ are placed at the two boundaries $(0,i(\tau_{0}+h(0))),(0,-i(\tau_{0}+h(0)))$ on the VWS. We want to compute Euclidean correlator of the type $\left<V(\omega_{1},\bar{\omega}_{1}) W (\omega_{2},\bar{\omega}_{2}) V(\omega_{3},\bar{\omega}_{3})\right>$. In the current set up, $(\omega_{i},\bar{\omega}_{i})_{i=1,2,3}$ are the following:
\begin{align}\label{omega}
\omega_{1} \equiv i\tau= i(\tau_0+h(0)), \omega_{2} = x, \omega_{3} \equiv -i\tau = -i(\tau_0+h(0)) \ .
\end{align}
While $\bar{\omega}_{i}$ take the corresponding complex conjugate values. We will then finally analytically continue to real time with ($\omega_1\rightarrow t+ i(\tau_0 +h(0)),\; \omega_2 = x \textrm{ and}\;  \omega_3 \rightarrow t -i(\tau_0 +h(0)))$. We will be interested in the large real $t$ limit and the observable we would like to compute is the normalized correlation function $\left(\frac{\left<VWV\right>}{\left<VV\right>\left<W\right>}\right)$. We calculate it by first mapping this to the CWS geometry and then further mapping it to the UHP geometry. On the UHP we use the doubling trick to compute the correlation functions. This whole set up is pictorially described in fig \ref{map plot}.

Using the map from VWS to CWS, we obtain: 
\begin{align}
\frac{\langle V(\omega_{1},\bar{\omega}_{1}) W (\omega_{2},\bar{\omega}_{2}) V(\omega_{3},\bar{\omega}_{3})\rangle \big|_{\rm VWS}}{\langle V(\omega_1,\bar{\omega}_1)V(\omega_3,\bar{\omega}_3)\rangle \langle W(\omega_2, \bar{\omega_2})\rangle\big|_{\rm VWS}} = \frac{\langle V(\zeta_{1},\bar{\zeta}_{1}) W (\zeta_{2},\bar{\zeta}_{2}) V(\zeta_{3},\bar{\zeta}_{3})\rangle \big|_{\rm CWS}}{\langle V(\zeta_1, \bar{\zeta}_1) V(\zeta_3, \bar{\zeta}_3)\rangle\langle W(\zeta_2, \bar{\zeta}_2)\rangle\big|_{\rm CWS}} \ .
\end{align}

Correlation functions on the CWS geometry can be mapped on to the UHP geometry by the standard map ($z = -ie^{\frac{\pi}{2\tau_0}\zeta}$). Thus the boundary points $\zeta_1$, $\zeta_2$ on the CWS are mapped to the points $z_1$, $z_2$ on the real line. As is well known, this bulk boundary three point function, has the same structure as that of a holomorphic four point function in the full plane and so the normalized three point function on the CWS is of the form
\begin{eqnarray}\label{CWS otoc}
&& \frac{\langle V(\zeta_{1},\bar{\zeta}_{1}) W (\zeta_{2},\bar{\zeta}_{2}) V(\zeta_{3},\bar{\zeta}_{3})\rangle \big|_{\rm CWS}}{\langle V(\zeta_1, \bar{\zeta}_1) V(\zeta_3, \bar{\zeta}_3)\rangle\langle W(\zeta_2, \bar{\zeta}_2)\rangle\big|_{\rm CWS}}= F(\eta(\zeta_i,\bar{\zeta}_i)), \; \; \\
&& \textrm{Where $\eta = \frac{(z_1-z_3)(z_2 -\bar{z}_2)}{(z_1-z_{2})(z_3-\bar{z}_2)}$ and}\; \; z_i =-ie^{\left(\frac{\pi}{2\tau_0} \zeta_i \right)} \ .
\end{eqnarray}
One can now compute the VWS correlation functions at the point $\omega_i$ by expanding $\zeta_i$ upto first order in $f$ expansion, $\zeta_i =\omega_i -f(\omega_i)$:
\begin{align}\label{cwsvws}
    &\frac{\langle V(\omega_{1},\bar{\omega}_{1}) W (\omega_{2},\bar{\omega}_{2}) V(\omega_{3},\bar{\omega}_{3})\rangle \big|_{\rm VWS}}{\langle V(\omega_1,\bar{\omega}_1)V(\omega_3,\bar{\omega}_3)\rangle \langle W(\omega_2, \bar{\omega_2})\rangle\big|_{\rm VWS}} = F(\eta(\omega_i,\bar{\omega}_i)) -\sum^3_{i=1}\Big(f(\omega_i)\frac{\partial}{\partial\omega_i} + f(\bar{\omega}_i)\frac{\partial}{\partial \bar{\omega}_i} \Big) F(\eta(\omega_i,\bar{\omega}_i)) \nonumber \\
    &= \Big[1 -2f(x)\partial_{x}-{\rm Re} f(\omega_{1})\partial_{\omega_{1}} - {\rm Re}f(\omega_{3})\partial_{\omega_{3}} - {\rm Re}\overline{f(\omega_{1})}\partial_{\bar{\omega}_{1}} - {\rm Re}\overline{f(\omega_{3})}\partial_{\bar{\omega}_{3}}\Big] F(\eta(\omega_i,\bar{\omega}_i))|_{\omega_{1}=i\tau_{0},\omega_{2}=x,\omega_{3}=-i\tau_{0}}  \ . 
\end{align}
Here in the second line, we have used the boundary condition of $g(\omega)$ as in (\ref{boundary cond g}) to write $F(\eta(\omega_i,\bar{\omega}_i))$ as a CWS bulk-boundary three point correlator where two boundary operators are located on the boundary of the strip $\pm i\tau_{0}$. After explicitly using (\ref{omega}) and the reflection symmetric property of $f$, we finally get
\begin{align}\label{final three point}
  &\frac{\langle V(\omega_{1},\bar{\omega}_{1}) W (\omega_{2},\bar{\omega}_{2}) V(\omega_{3},\bar{\omega}_{3})\rangle \big|_{\rm VWS}}{\langle V(\omega_1,\bar{\omega}_1)V(\omega_3,\bar{\omega}_3)\rangle \langle W(\omega_2, \bar{\omega_2})\rangle\big|_{\rm VWS}}= \left[ 1 - 2f(x)\partial_{x}\right]]F(\eta(\omega_i,\bar{\omega}_i))|_{\omega_{1}=i\tau_{0},\omega_{2}=x,\omega_{3}=-i\tau_{0}} \ . 
\end{align}

Hence the central object of the three point bulk-boundary OTOC computation in this inhomogeneous quench process, is the quantity $F(\eta(\omega_i,\bar{\omega}_i))$. In other words, $F$ determines the three point OTOC for global or homogeneous quench which is described purely by OTOC in CWS. The amount of inhomogeneity in the correlators at VWS is injected through the $f(x)$. As we see from (\ref{final three point}), in this special bulk-boundary setting, only spatial inhomogeneity enters in the first order correction.

Based on (\ref{final three point}), we can already anticipate the effect of inhomogeneity. For example, suppose $F(x, t) \sim e^{\lambda(t-x/v_{\rm B})}$, then inhomogeneity affects the butterfly velocity such that: 
\begin{eqnarray} \label{vb_inho}
v_{B}' = v_{B}\left(1 +2\frac{f(x)}{x}\right) \ . 
\end{eqnarray}
Let us discuss the implications of the above result. First, on physical grounds, we expect that the butterfly velocity provides us with the maximum speed at which information can propagate in the system,\footnote{Note that the Lieb-Robinson bound, for a system, is state-independent as it is defined in terms of the $L_{\infty}$-norm of operator commutators.} and therefore for any state, $v_{\rm B} \le 1$. Thus, while for $f(x)/x <0$, this is trivially satisfied, it is not so when $f(x)/x >0$. For the latter, when the homogeneous quench yields $v_{\rm B}=1$, $v_{\rm B}'>1$ and therefore violates a causality bound.\footnote{Recall that, even though the initial state is not Lorentz-invariant, it is nevertheless evolving with a CFT Hamiltonian after the quench. At sufficiently large times, the information propagation should be constrained by how fast information can propagate with the CFT-Hamiltonian. Thus, for a Lorentz-invariant CFT, we expect that the butterfly velocity should be upper bounded by the speed of light.} On the other hand, {\it a priori} there is no constraint on the sign of $f(x)$. Therefore, one compelling possibility is that this violation is a consequence of the leading order perturbative analyses, that may restore causality if higher order effects in $f(x)$ are kept and perhaps resummed. Later, we will discuss explicit examples in which this possibility is further evidenced. Let us now discuss the homogeneous quench case in more details.

\subsection{A Non-trivial $3$-point  $b$-OTOC: Homogeneous Quench}\label{sec3.2}

The homogeneous quench limit can be easily obtained by setting $f(x)=0$, in the discussion above. This yields, using equation (\ref{final three point}), the desired $3$-point correlator in terms of the function $F(\eta)$. Before discussing the $3$-point $b$-OTOC, let us begin with lower point correlators.

The basic features are nicely summarized in \cite{Calabrese:2016xau}, which we will heavily draw on in the subsequent discussion. First, an $n$-point function of primary, scalar operators, denoted by $\Phi_i(\omega_i)$, on the CWS can be mapped to the UHP. Recall that $\omega = x + i \tau$. Subsequently, to obtain the real-time correlator, we analytically continue: $\tau \to \tau_0 + i t $ and take the limit $\{t, x_{ij}\} \gg \tau_0$ to obtain the asymptotic time-dependence. It is straightforward to obtain the large time behaviour of a one-point function of a scalar, primary, of dimension $\Delta_\Phi$: $\langle \Phi(t) \rangle \approx (\pi/2\tau_0)^{\Delta_\Phi} e^{- \Delta_\Phi \pi t/ (2\tau_0)} $. It defines a relaxation time for the corresponding operator: $t_{\rm rel} \sim 2\tau_0/(\Delta_\Phi \pi)$. This already indicates that $\tau_0$ sets a dissipation-scale in the dynamics.

A more direct and explicit understanding is given by the one-point function of the stress-tensor. On the UHP, $\langle T_{\mu\nu} \rangle_{\rm UHP} =0$ and therefore on the CWS, the sole contribution comes from the Schwarzian piece of the conformal transformation. This yields: $\langle T_{\mu\nu} \rangle_{\rm CWS} = \pi c /(6 (4\tau_0)^2)$, where $c$ is the central charge. Comparing this with a thermal one-point function, we can read-off an effective temperature $\beta_{\rm eff} = 4\tau_0$. Similarly, two-point correlators can be also explicitly obtained in the asymptotic time limit. Using the results in \cite{Calabrese:2007rg}, it is easy to see that the connected equal-time correlator between two scalar primary operators, in this limit, is given by: 
\begin{eqnarray}\label{2-pointhomo}
&& \left \langle\Phi(\ell, t) \Phi(0,t) \right \rangle_{\rm conn} \equiv \left \langle\Phi(\ell, t) \Phi(0,t) \right \rangle - \left \langle \Phi(t) \right \rangle^2 \sim \left( \frac{\pi}{2\tau_0}\right)^{2\Delta_\Phi} e^{- \frac{\pi \ell \Delta_\Phi}{2\tau_0}} \ , \quad t > \frac{\ell}{2} \ , \\
&&  \left \langle\Phi(\ell, t) \Phi(0,t) \right \rangle_{\rm conn} \equiv \left \langle\Phi(\ell, t) \Phi(0,t) \right \rangle - \left \langle \Phi(t) \right \rangle^2 \sim 0 \ , \quad t < \frac{\ell}{2} \ ,
\end{eqnarray}
at the leading order.\footnote{Evidently, the correlator does not vanish identically, but the vanishing value implies that all contributions are exponentially suppressed, compare to $e^{- \frac{\pi \ell \Delta_\Phi}{2\tau_0}}$, which itself is also small in the $\ell \gg \tau_0$ limit.} The behaviour in (\ref{2-pointhomo}) hints the existence of an approximate light-cone at $t=\ell/2$.

This effective light-cone is further supported by the dynamical behaviour of entanglement entropy of a line-segment (denoted by $A$) of length $\ell$\cite{Calabrese:2006rx}: 
\begin{eqnarray} \label{entrophomo}
&& S_A = \frac{c}{3} \log\tau_0 + \frac{\pi c t}{6\tau_0} \ , \quad t < \frac{\ell}{2} \ , \\
&& S_A = \frac{c}{3} \log\tau_0 + \frac{\pi c \ell}{12\tau_0} \ , \quad t > \frac{\ell}{2} \ , 
\end{eqnarray}
that clearly demarcates a linearly growing behaviour in time, from a plateaux. This behaviour can be easily understood from a ballistic propagation of the quasi-particles produced by the initial state, with a propagation velocity $v_{\rm B}=1$. Dynamical behaviours in both (\ref{2-pointhomo}) and (\ref{entrophomo}) are identical to that of a thermal state of temperature $\beta_{\rm eff} = 4\tau_0$. Furthermore, as is explicitly shown in \cite{Calabrese:2006rx}, the reduced density matrix of a sub-system of length $\ell$, after sufficiently long time, becomes exponentially close to a thermal density matrix, with the same $\beta_{\rm eff}$. The arguments above generalize for an $n$-point function. The lesson, therefore, is that at the large time limit, global (homogeneous) critical quench dynamics can be described by an effective thermal physics.

Let us now consider the $3$-point $b$-OTOC in detail. Here, $g(\omega) = \omega$ in figure \ref{map plot}, we only need the map of the CWS to the UHP. On the CWS, the operators are placed at $V(+\tau_0, 0)$, $V(-\tau_0,0)$, $W(0,x)$. As is already mentioned in the previous section, we will place the operators at $\omega_{1,3}= \pm i \tau_0$ and $\omega_2=x$, and subsequently analytically continue the result to $\omega_1 \to t + i \tau_0$ and $\omega_3 \to t - i \tau_0$. Keeping track of this analytic continuation, the points $\omega_{1,2,3}$ are mapped to $z_1 = -i e^{b (t+i\epsilon_1)}$, $z_3 = -i e^{b(t+i\epsilon_2)}$ and $z_2 = -i e^{b (x+ i\epsilon_0)}$, with $b=\pi/(2\tau_0)$. Note that, we have included $i\epsilon_0$ in the location of the $W$ operator, which will be used to define the time-ordering. Furthermore, $\epsilon_1=\tau_0$ and $\epsilon_2 = -\tau_0$ ensures that, at $t=0$, $z_{1,3}$ are points on the ${\rm Im}(z)=0$ boundary of the UHP.\footnote{In principle, we can keep $\epsilon_{1,2}$ independent of $\tau_0$. However, in doing so, will map the $V$ operators away from the UHP-boundary. This is also an interesting correlator, however, since the $V$ operators move away from the UHP-boundary, they will generically be given by a six-point holomorphic function in the entire complex-plane. Constraining the $V$ operators on the UHP-boundary provides us with the simplest non-trivial OTOC for the system\cite{Das:2019tga}.} There are two inequivalent operator orderings in this set up: (i) $\epsilon_0 > \epsilon_1>\epsilon_2$, for which one obtains a Time-Order Correlator and (ii) $\epsilon_1 > \epsilon_0 > \epsilon_2$ , for which one obtains an Out-of-Time-Ordered Correlator. We will focus on the OTOC.

With these assignments, the invariant cross-ratio in (\ref{CWS otoc}) can be calculated, the details are given in equation (\ref{etacal}). As described in equation (\ref{z limit}), in the limit $t\to 0$ as well as $t \gg x$, $\eta \to 0$ from opposite directions, but for the OTOC-configuration, {\it i.e.}~when $\epsilon_1 > \epsilon_0>\epsilon_2$, $\eta$ can be larger than one. In the complex $\eta$-plane, one therefore moves to the second Riemann sheet, while crossing a branch-cut running from unity to infinity. This is the essential kinematic aspect that distinguishes between a TOC and an OTOC\cite{Roberts:2014ifa}. The dynamical information is contained further in the function $F(\eta)$, for a given CFT.

An interesting class of CFTs in which this can be explicitly calculated is the large $c$ CFTs, presumably with a Holographic dual. In this case, one takes the limit that $h_v$ is fixed and large while $h_w/c$ is fixed and small, in which case the explicit result is analytically known\cite{Fitzpatrick:2014vua}.\footnote{For more explicit details, see appendix \ref{A}, equation (\ref{Idblock}). Also, note that, as pointed in \cite{Roberts:2014ifa}, the final result appears to hold for $h_w\gg h_v\gg 1$. } The complete answer for the $3$-point $b$-OTOC takes the following form:
\begin{eqnarray}\label{otoexplicit}
 F(t) &\approx& 1 - \frac{24\pi h_W h_V}{\epsilon^{*}_{12}}  {\rm exp} \left[\frac{2\pi}{\beta_{\rm eff}}(t-t_*- x) \right]  \implies \lambda_{\rm L} = \frac{2\pi}{\beta_{\rm eff}} \ ,  v_{\rm B} = 1\ , \quad t < t_* +x \ ,\\
&\approx&  \left( \frac{\epsilon^{*}_{12}}{12 \pi h_W} \right)^{2 h_V} e^{- 2 h_V \lambda_{\rm L} (t-t_*-x)} \ , \quad t > t_* + x \ ,
\end{eqnarray}
where, the scrambling time $t_* = \beta_{\rm eff}/(2\pi) \log c$ and $\epsilon_{12}=i(e^{i\epsilon_{1}}-e^{i\epsilon_{2}})$. Clearly, $t_*$ is parametrically large compared to the dissipation scale, set by $\beta_{\rm eff}$. The corresponding exponential growth of the OTOC are characterized by a ``maximal" Lyapunov $\lambda_{\rm L} = (2\pi) /\beta_{\rm eff}$\footnote{This result is also obtained in \cite{ddas}.} and a butterfly velocity $v_{\rm B}=1$. The light-cone observed in (\ref{2-pointhomo}) matches exactly with this velocity. Finally, note that the result in (\ref{otoexplicit}) holds for any finite $\epsilon_{12}$, which works as a regulator and cuts off arbitrarily high energy modes. Equivalently, one can smear the operators $V$ over a Lorentzian time-scale $\Delta \tau$, such that each infinitesimal $\epsilon_{12}\to \Delta \tau$, in the formulae above\cite{Roberts:2014ifa}. Thus, all conclusions above hold for any finite value of $\tau_0$. 

The discussion of this section is consistent with, and reinforces, the observation made in \cite{Calabrese:2006rx}, that the CC state at late times behaves as a thermal state. Given this fact, it is natural to wonder whether some version of the chaos bound derived in \cite{Maldacena:2015waa}, exists for OTOC's in the CC state. We will sketch here a proof of the same within the framework of the 3pt $b$-OTOC, discussed in this section. The key mathematical content of the proof of \cite{Maldacena:2015waa} is the existence of an upper bound on the rate of growth of $f$, in particular an upper bound on $\frac{1}{1-f(t)}|\frac{df(t)}{dt}|$, given in equation (4.1) of that paper. This bound exists when the following criteria are satisfied: (1)  the function $f(t+i\tau)$ is analytic inside a semi infinite strip  $t>0$ and $-\frac{\beta}{4}\leq \tau\leq \frac{\beta}{4}$, (2) $f(t+i\tau)$ is real for $\tau=0$ and (3) $|f(t+i\tau)|\leq 1$, in this strip. The function was essentially $f(t+t_o) = \frac{\textrm{Tr}[yVyW(t+t_0)yVyW(t+t_0)]}{\textrm{Tr}[y^2Vy^2V]\textrm{Tr}[y^2W(t+t_0)y^2W(t+t_0)] +\epsilon}$.  Where $y= e^{-\frac{\beta}{4}H}$, while $\epsilon$ and $t_0$ are defined so that condition 3 is met. Condition (1) and (2), is automatically satisfied by the definition of $f$.  For the CC state, the natural candidate would be $f(t) =\frac{<V(0-i\tau_0)W(t)V(0+i\tau_0)>}{<VV><W> +\epsilon}$, where the correlation functions are computed in the bCFT. By the doubling trick, this would behave like a four point function in a bulk CFT, and so we expect that arguments similar to what went into showing that condition (3) is met would still hold in this case. In particular at late times when the CC state behaves like the thermal state, the factorization property of the correlation function would also hold.  And so, as before with a suitably defined $\epsilon$, the corresponding $f$ satisfies condition(3). Condition (1) and (2) is again satisfied by the definition of $f$. The bound then follows along the same lines as given in section 4 of \cite{Maldacena:2015waa}.

\subsection{Inhomogeneous quench: A Special Limit $\tau_0\to0$}\label{sec3.3}

We will now introduce an inhomogeneous initial state, characterized by $\{\tau_0, h(x)\}$, in the limit $\tau_0 \gg {\rm sup}[h(x)]$, for $x \in [-\infty, \infty]$. Here, we will consider the limit $\tau_0\to 0$, while maintaining $\tau_0 \gg {\rm sup}[h(x)]$. Suppose, we choose the following function $h(x)$:
\begin{align}\label{choice f1}
h(x) = a \frac{e^{bx}}{1+  e^{bx}} \; \; \; \text{with} \; \; \; a, b>0 \ , \; \; \; \text{such that} \quad f(x) = \frac{a}{\tau_{0} b}\log\left(1+  e^{bx} \right) \ .
\end{align}
Clearly, $h[-\infty] =0$ and $h[+\infty] = a$. We need to impose $\tau_0 \gg {\rm sup}[h(x)] = a$, and the scale of variation of $h(x)$ is determined by $x_{\rm var}\sim 1/b$. On the other hand, the function $f(x) \approx e^{bx}\frac{a}{\tau_0 b}$ in the range: $b x \ll 0$. Therefore, we can place the bulk operator at a sufficiently negative value of $x$, satisfying: $  - \infty \ll bx \ll 0$, such the above approximation holds and also the function $h(x)$ is not vanishingly small. In the other regime: $b x \gg 0$, $f(x) \approx \frac{a}{\tau_0 }x$.

{\bf $2$-point correlation function}: Let us begin with a discussion on the $2$-point correlator, as we did for the homogeneous quench. The disconnected correlation function can be calculated easily, see {\it e.g.}~\cite{Sotiriadis:2008ila}. Without any loss of generality, let us discuss a specific example: $h(x)=ae^{\frac{\pi x}{2\tau_0}}\left( be^{\frac{\pi x}{2\tau_0}}+1\right)^{-1}$. The equal time two point correlation function in VWS, in the limit $\tau_0 \rightarrow 0$, is given by:
\begin{align}
    &C(x_1,x_2)\nonumber \\
    &= e^{-\frac{\Delta_{\Phi}\pi(x_1-x_2)}{2\tau_0}}\Bigl(1+ \frac{a\Delta_{\Phi}}{2b\tau_0}\Bigl[\log \left(be^{\frac{\pi(x_1-t)}{2\tau_0}}+1 \right)+\log \left(be^{\frac{\pi(x_1+t)}{2\tau_0}}+1 \right) - \log \left(be^{\frac{\pi(x_2-t)}{2\tau_0}}+1 \right) - \log \left(be^{\frac{\pi(x_2+t)}{2\tau_0}}+1 \right) \Bigr]\nonumber\\
 &-\frac{a\Delta_{\Phi}}{2\tau_0}\left[\frac{e^{\frac{\pi(x_{1}+t)}{2\tau_0}}}{be^{\frac{\pi(x_{1}+t)}{2\tau_0}}+1}+\frac{e^{\frac{\pi(x_1+t)}{2\tau_0}}}{be^{\frac{\pi(x_1+t)}{2\tau_0}}+1}+\frac{e^{\frac{\pi(x_1-t)}{2\tau_0}}}{be^{\frac{\pi(x_1-t)}{2\tau_0}}+1}+\frac{e^{\frac{\pi(x_2+t)}{2\tau_0}}}{be^{\frac{\pi(x_2+t)}{2\tau_0}}+1}+\frac{e^{\frac{\pi(x_2-t)}{2\tau_0}}}{be^{\frac{\pi(x_2-t)}{2\tau_0}}+1}\right]\Bigr) \nonumber\\
   & \approx e^{-\frac{\Delta_{\Phi}\pi(x_1-x_2)}{2\tau_0}} \left [1-\frac{a\Delta_{\Phi}}{\tau_0} \left (e^{\frac{\pi(x_2-t)}{2\tau_0}}+e^{\frac{\pi(x_2+t)}{2\tau_0}} \right ) \right] , \; \text{when}\; t>\frac{(x_1-x_2)}{2}.
    \end{align}
In the second line of the above equation we have considered $be^{\frac{\pi(x_i-t)}{2\tau_0}}\ll 1$, for $i=1,2$.

Thus for, $t \gg x_2$, the above equation yields:
\begin{align}
    C(x_1,x_2)\approx e^{-\frac{\Delta_{\Phi}\pi(x_1-x_2)}{2\tau_0}}-\frac{a\Delta_{\Phi}}{2\tau_0} \left (e^{\frac{\pi}{\tau_0}(x_2+t-\chi(x_1-x_2))} \right) \ . 
    \end{align}
Therefore, due to the sub-leading contribution, there is a growth in `$t$' at late times. Similarly, for $t<\frac{(x_1-x_2)}{2}$ and $be^{\frac{\pi(x-t)}{2\tau_0}}\ll 1$, the correlation function is:
\begin{align}
    C(x_1,x_2)&\approx e^{-\frac{\Delta_{\Phi}\pi t}{\tau_0}} \left [1-\frac{a\Delta_{\Phi}}{\tau_0} \left (e^{\frac{\pi(x_1-t)}{2\tau_0}}+e^{\frac{\pi(x_2-t)}{2\tau_0}} \right ) \right]\nonumber\\
    &\approx e^{-\frac{\Delta_{\Phi}\pi t}{\tau_0}}-\frac{a\Delta_{\Phi}}{\tau_0} \left(e^{\frac{\pi}{2\tau_0}(x_1-(2\Delta_{\Phi}+1))t}+e^{\frac{\pi}{2\tau_0}(x_2-(2\Delta_{\Phi}+1))t} \right) \ . 
    \end{align}
Therefore, for $t\ll x_1,x_2$, there is an overall decay at early times, even after considering sub-leading contributions. Using these, we calculate the connected correlation function $\langle \phi \phi \rangle-(\langle\phi\rangle)^{2}$. This yields a vanishing answer for $t< \frac{(x_1-x_2)}{2}$. Hence, the equal-time $2$-point function sees the same light-cone as it does for the homogeneous quench. Perhaps more importantly, the light-cone has no memory of the initial inhomogeneous state and coincides with the relativistic CFT light-cone. We will see momentarily that higher point functions, specially the OTOCs encode a different light-cone altogether.

{\bf Higher-point functions \& OTOCs}: We will now discuss specific examples, in which we compute the 3-point $b$-OTOC in explicit details. The simplest example is large-$c$ CFTs, in which, as before, we take the $W$ operators to be heavy and the $V$ operator to be light. Using the explicit result for the identity block in equation (\ref{Idblock}), and substituting in (\ref{cwsvws}), one obtains:
\begin{align}
&\frac{\langle V(\omega_{1},\bar{\omega}_{1}) W (\omega_{2},\bar{\omega}_{2}) V(\omega_{3},\bar{\omega}_{3})\rangle \big|_{\rm VWS}}{\langle V(\omega_1,\bar{\omega}_1)V(\omega_3,\bar{\omega}_3)\rangle \langle W(\omega_2, \bar{\omega_2})\rangle\big|_{\rm VWS}} \nonumber\\
& \approx 1 - \frac{24\pi h_{w}h_{v}}{\epsilon_{12}^{*}} e^{\frac{\pi}{2\tau_{0}} (t -t_{*} -x)}  - \frac{24\pi^2 a h_{v}h_{w}}{\tau_{0}^2  \epsilon_{12}^{*}b} e^{ \frac{\pi}{2\tau_{0}} \left(t - t_{*} - \left(1-\frac{2\tau_{0}}{\pi}b \right) x \right)}, \; \text{when}\; t<t_{*} \ , \\
& \approx \left(\frac{\epsilon^{*}_{12}}{12\pi h_{w}}\right)^{2h_{v}}e^{-\frac{\pi h_{v}}{\tau_{0}}(t-t_{*}-x)} - \frac{24\pi^2 ah_{w}h_{v}}{\tau_{0}^2\epsilon_{12}^{*} b}\left(\frac{\epsilon^{*}_{12}}{12\pi h_{w}}\right)^{2h_{v}+1}e^{-\frac{\pi h_{v}}{\tau_{0}} \left(t - t_{*}- \left(1 + \frac{\tau_{0} b}{\pi h_{v}}\right )x \right)},\; \text{when}\; t>t_{*} \ , 
\end{align}
where $t_* = (2\tau_0/\pi) \log c$ and we have kept only leading order terms in an $(1/c)$-expansion.

Below the scrambling time, the corrections due to inhomogeneity can be exponentiated\footnote{This exponentiation is valid for $\log 2 + \log(a/\tau_0) + bx \ll 0$, which is guaranteed since $a\ll \tau_0$ and $b x \ll 0$. } to yield:
\begin{eqnarray} \label{otocinhomlargec}
F(t) & = & 1 - \frac{24 \pi h_w h_v}{c \epsilon_{12}^*} {\rm exp}\left[ \frac{\pi}{2\tau_0} \left( t - x \right) - \log\left(1+\frac{\pi a}{\tau_0^2 b} e^{bx}\right) \right] , \\
& \approx & 1 - \frac{24 \pi h_w h_v}{c \epsilon_{12}^*} e^{\frac{a\pi}{b\tau_0^2}} {\rm exp}\left[ \frac{\pi}{2\tau_0} \left( t - \left(1 - \frac{2 a}{\tau_0} \right) x  \right) \right]  .
\end{eqnarray}
From the second line above, we can read off the corresponding Lyapunov exponent as well as the butterfly velocity: $\lambda_{\rm L} = \pi / (2\tau_0)$ and $v_{\rm B} = 1 + (2a)/\tau_0$, for $\tau_0 \gg a$. While $\lambda_{\rm L}$ receives no correction, there is a change in the light-cone structure since $v_{\rm B}$ is modified. Correspondingly, the scrambling time is modified in (\ref{otocinhomlargec}), with an order one change: $t_* \sim \left(1/\lambda_{\rm L} \right) \log \left(c e^{-a\pi /\tau_0^2b} \right)$. As we now explicitly observe: $v_{\rm B} > 1$ when $a>0$.

To better understand this, consider the  example: $h(x) = a \Theta (x-a_1)$, with $a \ll \tau_0$ and $a_1$ is arbitrary. This yields: $f(x) = a (x-a_1)\Theta(x-a_1)$. Repeating the same calculation as above, we get
\begin{align}\label{step three point}
  &\frac{\langle V(\omega_{1},\bar{\omega}_{1}) W (\omega_{2},\bar{\omega}_{2}) V(\omega_{3},\bar{\omega}_{3})\rangle \big|_{\rm VWS}}{\langle V(\omega_1,\bar{\omega}_1)V(\omega_3,\bar{\omega}_3)\rangle \langle W(\omega_2, \bar{\omega_2})\rangle\big|_{\rm VWS}} \nonumber \\
 & \approx 1 - \frac{24\pi h_{w}h_{v}}{\epsilon_{12}^{*}} e^{\frac{\pi}{2\tau_{0}}(t-t_{*}-x)}- a(x-a_{1})\theta(x-a_{1}) \frac{24\pi^{2} h_{v}h_{w}}{\tau_{0}\epsilon_{12}^{*}}e^{\frac{\pi}{2\tau_{0}}(t-t^{*}-x)} \text{when}\; t< t_{*} \ ,  \nonumber \\
& \approx \left[ \left(\frac{\epsilon^{*}_{12}}{12\pi h_{w}}\right)^{2h_{v}}-a \frac{24\pi^{2} h_{w}h_{v}(x-a_{1})\theta(x-a_{1})}{\tau_{0}\epsilon_{12}^{*}}\left(\frac{\epsilon^{*}_{12}}{12\pi h_{w}}\right)^{2h_{v}+1}\right]e^{-\frac{\pi h_{v}}{\tau_{0}}(t-t_{*}-x)},\; \text{when}\; t>t_{*} \ . 
\end{align}
Re-arranging the above expressions in (\ref{step three point}), we can read-off the Lyapunov exponent and the butterfly velocities as: $\lambda_{\rm L} = \frac{\pi}{2\tau_0}$, $v_{\rm B} (x) =  1 + 2 a \Theta(x-a_1)$. A similar exercise can be carried out when $h(x)$ is the bump function: $h(x) = a(\Theta(x-a_{1})-\Theta(x-a_{2}))$ with $a \ll \tau_0$ and $a_2> a_1$. The Lyapunov exponent remains unchanged, as above. The butterfly velocity, on the other hand, is given by $v_{\rm B} = 1 + 2 a \Theta(x-a_1) + 2a \Theta(x-a_2)$.

It is natural to wonder --- since the inhomogeneity is introduced by a simple step function --- why the Lyapunov exponent only sees $\tau_0$ and not $(\tau_0 + a)$. Indeed, physically, $\lambda_{\rm L} = \frac{\pi}{2(\tau_0+a)}$ in the limit $a_1 \to -\infty$. It is easy to check, however, that in a perturbative expansion in $(a/\tau_0)$, the effect of $\lambda_{\rm L} = \pi/(\tau_0 +a)$ is only visible at $\O(a/\tau_0^2)$, which is outside the regime of validity of our expansion. In fact, it is non-perturbative in the limit $a\ll \tau_0 \ll 1$. In the same limit, we also expect $v_{\rm B}=1$, since this is arbitrarily close to a homogeneous quench. It is suggestive that a similar non-perturbative effect may underlie how this can be reconciled with the perturbative answer $v_{\rm B}(x) = 1 + 2 a \Theta(x-a_1)$.

\numberwithin{equation}{section}
\section{A 4-pt bulk OTOC: large $\tau_0 $ limit} \label{sec4}

In the previous section, we have seen that a three point $b$-OTOC already captures a change in the light-cone structure and in the scrambling time, at the leading order perturbative calculation in inhomogeneity, for a large-$c$ critical quench. It is further clear from equation (\ref{final three point}) that the temporal part remains unaltered in the $3$-point $b$-OTOC. We will now show, by analyzing higher point functions, that the Lyapunov exponent also receives corrections due to inhomogeneity of the initial state. Our arguments will rely on general structure of higher point functions in the limit where bulk operators approach close to each other, compared to their proximity to the boundary. This is precisely the $\tau_0 \gg \text{dist}(\omega_{i},\omega_{j})$ limit, which is complementary to the analyses above. Here $\text{dist}(\omega_{i},\omega_{j})$ refers to the distance between any two points where the probe operators are located. In other words, this limit tells the distances between probe operators must be very small compared to the distance from the boundary.

Let us consider two scalar operators $V$ of dimension $h_{V}(=\bar{h}_{V})$, placed at $(0,i\tau)$ and two other scalar operators $W$ of dimension $h_{W}$, placed at $(x,0)$ on the VWS. Consider the four-point correlator of the type $\left \langle V(\omega_{1},\bar{\omega}_{1}) W (\omega_{2},\bar{\omega}_{2}) V(\omega_{3},\bar{\omega}_{3}) W (\omega_{4},\bar{\omega}_{4}) \right\rangle$. In this case, we have:
\begin{align}\label{s13}
\omega_{1}= \omega_{3} = i\tau \ , \quad \omega_{2}= \omega_{4} = x; 
\end{align}
and $\bar{\omega}_{i} \rightarrow$ corresponding complex conjugates. We compute the normalized correlation function $\frac{\langle VWVW\rangle}{\langle VV\rangle \langle WW\rangle}$ by following the same (VWS $\rightarrow $ CWS $\rightarrow$ UHP) mapping scheme to map the problem in UHP, as demonstrated in figure \ref{map plot}.

The bulk four point function on the UHP has the same structure as a holomorphic eight point function in the full plane. However, in the limit $x,\tau \ll \tau_{0}$, we can ignore the boundary effect. However, we can not get any non-trivial OTOC in this limit from lower point(e.g three point) bulk correlators. This implies that in the UHP, the eight point function is factorized into a product of two $4$-point functions of bulk and it's image points {\it i.e}~$F(z_{i},\bar{z}_{i}) \approx g(z_{i})g(\bar{z}_{i})$. Here $z_{i} = e^{\frac{\pi\omega_i}{2\tau_{0}}}$, as we have mentioned earlier. However, notice that this map is exactly same as the vacuum to thermal(or plane($\omega$) to cylinder($z$)) map with temperature $\beta$ {\it i.e.}~$\omega \rightarrow z = e^{\frac{2\pi z}{\beta}}$\cite{Calabrese:2006rx}\footnote{ One arrives at the same conclusions in large real time(with finite or small $\tau_{0}$) limit for equal time or time ordered correlators, as argued in \cite{Calabrese:2006rx}.}. Hence in this large $\tau_{0}$ limit, the decoupled four point OTOCs give the same answer as that in a thermal ensemble with the identification $\beta=4\tau_{0}$, adding further evidence to our previous discussion based on the $3$-point $b$-OTOC. The $\tau_0\gg (x,\tau)$ limit is tractable in both homogeneous as well as inhomogeneous quench.

Proceeding as before, we find the normalised correlator on VWS:
\begin{align}
    &\frac{\langle V(\omega_{1},\bar{\omega}_{1}) W (\omega_{2},\bar{\omega}_{2}) V(\omega_{3},\bar{\omega}_{3}) W (\omega_{4},\bar{\omega}_{4})\rangle \big|_{\rm VWS}}{\langle V(\omega_1,\bar{\omega}_1)V(\omega_3,\bar{\omega}_3)\rangle \langle W(\omega_2, \bar{\omega_2})W(\omega_4, \bar{\omega_4})\rangle\big|_{\rm VWS}}\nonumber\\&= F(\eta_{1}(\omega_i,\bar{\omega}_i),\eta_{2}(\omega_i,\bar{\omega}_i)) -\sum^3_{i=1}\Big(f(\omega_i)\frac{\partial}{\partial\omega_i} + f(\bar{\omega}_i)\frac{\partial}{\partial \bar{\omega}_i} \Big) F(\eta_{1}(\omega_i,\bar{\omega}_i),\eta_{2}(\omega_i,\bar{\omega}_i)) \nonumber \\
    &= \Big[1 -4f(x)\partial_{x}+2i(f(i\tau)-f(-i\tau))\partial_{\tau}\Big]F(\eta_{1}(\omega_i,\bar{\omega}_i),\eta_{2}(\omega_i,\bar{\omega}_i))|_{\omega_{1}=\omega_{3}=i\tau,\omega_{2}=\omega_{4}=x} \ . 
\end{align}
After carrying out the analytic continuation using $i\epsilon$ prescription from $i\tau \rightarrow t+i\epsilon$, we get:
\begin{align}\label{VWS final 4pt}
    &\frac{\langle V(\omega_{1},\bar{\omega}_{1}) W (\omega_{2},\bar{\omega}_{2}) V(\omega_{3},\bar{\omega}_{3})W (\omega_{4},\bar{\omega}_{4})\rangle \big|_{\rm VWS}}{\langle V(\omega_1,\bar{\omega}_1)V(\omega_3,\bar{\omega}_3)\rangle \langle W(\omega_2, \bar{\omega_2})W(\omega_4, \bar{\omega_4})\rangle\big|_{\rm VWS}}\nonumber\\
    &= \Big[1 -4f(x)\partial_{x}-2(f(t+i\epsilon_{1})-f(-t-i\epsilon_{3}))\partial_{t}\Big]F(\eta_{1}(\omega_i,\bar{\omega}_i),\eta_{2}(\omega_i,\bar{\omega}_i)) \ ,
\end{align}
which has manifest temporal derivatives and therefore will affect the time-dependence. In the limit $\tau_{0} \gg x,\tau$, $F(\eta_{1}(\omega_i,\bar{\omega}_i),\eta_{2}(\omega_i,\bar{\omega}_i)) \rightarrow g(z_{i})g(\bar{z}_{i})$, where $g(z_{i}) = \langle V(z_{1}) W (z_{2}) V(z_{3})W (z_{4})\rangle $. As we argued before, this $g$ can be computed from the thermal four point OTOC in CFT$_{2}$. Hence, the correction due to inhomogeneity comes purely from the temporal and spatial derivatives in (\ref{VWS final 4pt}).

As before, based on (\ref{VWS final 4pt}) we can already anticipate the effect of inhomogeneity on the corresponding Lyapunov exponent. First of all, for an monotonically increasing function $f(x) (\text{and} f(t))$ taking a schematic functional dependence $F(x,t) \sim e^{\lambda_{\rm L}(t-x/v_{\rm B})}$, one can have
\begin{eqnarray}
\left( 1-4f(x)\partial_{x} - 2f(t)\partial_{t} \right)e^{\lambda_{\rm L} \left(t-x/v_{\rm B} \right )} = e^{\lambda'_{\rm L} \left(t-x/v'_{\rm B} \right)} \ .
\end{eqnarray}
For $2\lambda_{\rm L}\left( 2f(x)/v_{\rm B} - f(t)\right) \ll 1$ and $f(x) \ll x, f(t) \ll t$, we obtain:
\begin{align}
\lambda'_{\rm L} = \lambda_{\rm L}\left(1-\frac{2f(t) - 2 f(-t)}{t}\right), \; v'_{\rm B} = v_{\rm B}\left(1 + \frac{4f(x)}{x} - \frac{2f(t)}{t}\right)
\end{align}
As before, the correction to butterfly velocity can exceed speed of light by an amount $\frac{4f(x)}{x}-\frac{2f(t)}{t} >0 \, \, (\text{for} \, \, t\gg x)$. Also it is easy to observe that for an even function $f(-t) = f(t)$, there is no correction to $\lambda_{\rm L}$ and $v'_{\rm B} = v_{\rm B}\left(1 + \frac{4f(x)}{x}\right)$. However for an odd function, $f(-t) = - f(t)$, we have $\lambda'_{\rm L} = \lambda_{\rm L}\left(1-\frac{4f(t)}{t}\right), \; v'_{\rm B} = v_{\rm B}\left(1 + \frac{4f(x)}{x} - \frac{4f(t)}{t}\right)$. Even in this case for $t \gg x$, $v'_{\rm B} > v_{\rm B}$. As we have argued before, this violation is likely to be visible only in the leading order perturbative answer. This is, nonetheless, measurable since an observer can certainly work within this approximation.

More generally, from (VWS final 4pt), it is clear that for any system that has a non-trivial time-dependence in the CWS correlator, the corresponding VWS correlator will have a universal exponential growth, provided by the external data $f(t) \sim e^{\kappa t}$, with a Lyapunov exponent set by $\kappa$. Moreover, the term $(1- f(t) \partial_t)$ naturally defines a time-scale $t_* \sim - \tau_0 \log\left({\rm sup}[f(t)]\right) \gg \tau_0$. Here, ${\rm sup}[f(t)]$ denotes the maximum modulus of the function $f(t)$, which by assumption is still a small number and $\tau_0$ appears on dimensional grounds. Therefore, we obtain a natural definition of scrambling time, without making any reference to large number of degrees of freedom. Later we will discuss an explicit example of this. More generally, it is clear from the analyses leading to (\ref{VWS final 4pt}) that higher point functions will also have a similar behaviour.

To further elucidate with explicit examples, we use (\ref{f(x)}) to obtain $f(x)$ in large $\tau_{0}$ limit. For analytical control, we choose: $h(x) = a\frac{e^{\pi x/\tau_{0}}}{1+de^{\pi x/\tau_{0}}}$, such that $(a/d) \ll \tau_0$. This yields:
\begin{align}\label{f(x) large tau}
f(z) &= a\ln\left(\frac{1}{d}\right)\frac{e^{\pi z/\tau_{0}}}{\pi(1-de^{\pi z/\tau_{0}})} - a\frac{z}{\tau_{0}}\frac{e^{\pi z/\tau_{0}}}{1-de^{\pi z/\tau_{0}}} \nonumber \\
&\approx a\frac{1}{\pi}\ln\left(\frac{1}{d}\right)\left(e^{\pi z/\tau_{0}} + de^{2\pi z/\tau_{0}}\right) -a\frac{1}{\tau_{0}}\left(ze^{\pi z/\tau_{0}} + dze^{2\pi z/\tau_{0}}\right); \; \forall \; z<\frac{\tau_{0}}{\pi}\ln\left(\frac{1}{d}\right) \ . 
\end{align}
Furthermore, in the $\tau_0 \gg (x,\tau)$ limit, $e^{2\pi z/\tau_0}$-term can be ignored by arranging sufficiently small $d$. With this, let us now discuss specific examples.

\subsection{Large $c$ CFT$_{2}$:}\label{sec4.1}

This is an interesting class of CFTs, for which the $4$-point CWS-correlator is given by a thermal correlator with an effective $\beta_{\rm eff} = 4\tau_0$. Using the identity block dominance, the corresponding thermal correlator is obtained to be:
\begin{align}
F(\eta_{1}(\omega_i,\bar{\omega}_i),\eta_{2}(\omega_i,\bar{\omega}_i)) &\sim g(\eta(\omega_{i}))g(\eta(\bar{\omega}_{i})) \sim \left(\frac{1}{1+\frac{24i\pi h_{w}}{c \epsilon_{12}^{*}\epsilon_{34}}e^{\frac{\pi}{2\tau_{0}}(t-x)}}\right)^{2h_{v}} \ ,
\end{align} 
which is very similar to the discussion in (\ref{Idblock}). Here $\epsilon_{1,2,3,4}$ specify the analytic continuations corresponding to the points $\omega_{1,2,3,4}$.

Using the above expression and the appropriate form of $f(x),f(t)$ as in (\ref{f(x) large tau}), the equation (\ref{VWS final 4pt}) yields:
\begin{align}
 &\frac{\langle V(\omega_{1},\bar{\omega}_{1}) W (\omega_{2},\bar{\omega}_{2}) V(\omega_{3},\bar{\omega}_{3})W (\omega_{4},\bar{\omega}_{4})\rangle \big|_{\rm VWS}}{\langle V(\omega_1,\bar{\omega}_1)V(\omega_3,\bar{\omega}_3)\rangle \langle W(\omega_2, \bar{\omega_2})W(\omega_4, \bar{\omega_4})\rangle\big|_{\rm VWS}}\nonumber\\
    &\approx 1-\frac{48i\pi h_{w}h_v}{c \epsilon_{12}^{*}\epsilon_{34}}e^{\frac{\pi}{2\tau_{0}}(t-x)} {-} a\frac{4}{\pi}\ln\left(\frac{1}{d}\right)\left(e^{\pi x/\tau_{0}} \right) \left(\frac{24h_{v}h_{w}\pi^{2}i}{c \tau_{0}\epsilon_{12}^{*}\epsilon_{34}}e^{\frac{\pi}{2\tau_{0}}(t-x)}\right) \nonumber \\
 &{+} a\frac{2}{\pi}\ln\left(\frac{1}{d}\right)\left(e^{\pi t/\tau_{0}} -(t \leftrightarrow -t)\right) \times \left( \frac{24h_{v}h_{w}\pi^{2}i}{c \tau_{0}\epsilon_{12}^{*}\epsilon_{34}}e^{\frac{\pi}{2\tau_{0}}(t-x)}\right), \; \text{when} \; t<t_{*} \\
 & \approx \left(c \frac{\epsilon_{12}^{*}\epsilon_{34}}{24i\pi h_{w}}\right)^{2h_{v}}e^{-\frac{\pi}{\tau_{0}}h_{v}(t-x)}\nonumber \\
 &- a\frac{4}{\pi}\ln\left(\frac{1}{d}\right)\left(e^{\pi x/\tau_{0}} \right) \left(\frac{\epsilon_{12}^{*}\epsilon_{34}}{24i\pi h_{w}}\right)^{2h_{v}+1} \frac{24\pi^{2}ih_{v}h_{w} c^{2h_v}}{\tau_{0}\epsilon_{12}^{*}\epsilon_{34}}e^{-\frac{\pi}{\tau_{0}}h_{v}(t-x)} + \nonumber \\
    &+a\frac{2}{\pi}\ln\left(\frac{1}{d}\right)\left(e^{\pi t/\tau_{0}} -(t \leftrightarrow -t)\right)\left(\frac{\epsilon_{12}^{*}\epsilon_{34}}{24i\pi h_{w}}\right)^{2h_{v}+1} \frac{24\pi^{2}ih_{v}h_{w}c^{2h_v}}{\tau_{0}\epsilon_{12}^{*}\epsilon_{34}}e^{-\frac{\pi}{\tau_{0}}h_{v}(t-x)}, \; \text{when} \; t>t_{*} \ .
\end{align}
Note that, here scrambling time  $t_* \sim \log\left( c e^{a/\tau_0\ln\left(\frac{1}{d}\right)}\right) $. Below $t_*$, the above sum over exponentials can be re-written as a single exponential term. This yields a modified Lyapunov exponent and butterfly velocity:
\begin{eqnarray} \label{correctioninho}
\lambda_{\rm L} = \frac{\pi}{2\tau_0} {- \frac{a\pi}{\tau_0^2}} \log\left( \frac{1}{d}\right) \ , \quad v_{\rm B} = 1{+\frac{2a}{\tau_0}} \log\left( \frac{1}{d}\right) \ .
\end{eqnarray}
As before, for $a>0$, $v_{\rm B}>1$. On the other hand, in this case $\lambda_{\rm L} < \frac{\pi}{2\tau_0}$ and therefore the $3$-point $b$-OTOC provides us with the maximum of the Lyapunov spectrum. For $a<0$, $v_{\rm B}<1$ and $\lambda_{\rm L}>\frac{\pi}{2\tau_0}$. In this case, the $4$-point OTOC defines the maximum of the Lyapunov-spectrum.

Several comments are in order: First, the similarity between a critical quench and a thermal state is suggestively universal, far outside the universality of $2$-point and $3$-point correlators in the CFT. The corresponding dynamical data $\{\lambda_{\rm L}, v_{\rm B}\}$ are independent of the specific operators, and universally depend only the quench data $\{\tau_0, h(x)\}$. This universality is already expected from (\ref{VWS final 4pt}) and we will later discuss other explicit examples as well.

Secondly, higher point functions are clearly more sensitive to the initial data. For example, as we have discussed above, the $2$-point correlator yields a light-cone co-incident with the one defined by the speed of light. This, however, is not the case for OTOCs. Instead, recovering a relativistic light-cone from OTOCs appear to be a non-trivial task that lies outside the leading order perturbative analyses that we have used. This suggests that such OTOCs encode more than universal CFT dynamics of low-point correlation functions. We will later discuss an interesting ramification of this with minimal models.

Third, from a Holographic perspective a {\it thermal-dynamics} of critical quenches resonates with an emergent thermal-like behaviour in other contexts, see {\it e.g.}~\cite{Das:2010yw, Kundu:2019ull}. In presence of inhomogeneity, this is, however, a mixed bag: For example, while the dynamical data of (\ref{correctioninho}) are thermal-like, the expectation value of the stress-tensor one-point function exhibits a non-diffusive wave-like propagation\cite{Sotiriadis:2008ila}. Nonetheless, the Holographic dual description can be approximated by a black hole geometry, as a result of gravitational collapse of an initial inhomogeneous data.

We will now discuss minimal models, as examples of super-integrable systems.

\subsection{Minimal models ($c<1$) and free fermions:} \label{sec4.2}

Let us now consider integrable 2D CFTs {\it e.g.}~Rational CFT and free Fermions, both of which have only an order one number of degrees of freedom. For Unitary minimal models, thermal OTOC has been studied extensively in \cite{Fan:2018ddo}, and we will use their results. At very large time, the corresponding OTOCs approach a universal constant value which is completely determined by modular S matrix. In this limit, we will also obtain trivial result for the CWS as well as the VWS OTOCs. In the regime $0<t-x\ll \tau_{0}$, however, the CWS OTOC shows a dynamical behaviour of the form: $e^{2\pi i}(h_{q_{1}}-h_{w}-h_{v})\left(1+ \gamma (t-x)^{2 \left(h_{q_{2}}-h_{q_{1}} \right)}\right)$, where  $h_{q_{1}}$ and $h_{q_{2}}$ are two smallest conformal dimensions of $WV$ fusion channel, and $\gamma$ is a numerical constant.

Using this behaviour we can now compute the corresponding VWS OTOC as follows:
\begin{align}
&\frac{\langle V(\omega_{1},\bar{\omega}_{1}) W (\omega_{2},\bar{\omega}_{2}) V(\omega_{3},\bar{\omega}_{3})W (\omega_{4},\bar{\omega}_{4})\rangle \big|_{\rm VWS}}{\langle V(\omega_1,\bar{\omega}_1)V(\omega_3,\bar{\omega}_3)\rangle \langle W(\omega_2, \bar{\omega_2})W(\omega_4, \bar{\omega_4})\rangle\big|_{\rm VWS}} \nonumber\\
& \sim e^{2\pi i} \left(h_{q_{1}}-h_{w}-h_{v})(1+ \gamma (t-x)^{2(h_{q_{2}}-h_{q_{1}})} \right) \nonumber\\
&+ \frac{8a\gamma}{\pi}\ln\left(\frac{1}{d}\right)e^{2\pi i}(h_{q_{1}}-h_{w}-h_{v}) (h_{q_{2}}-h_{q_{1}}) (t-x)^{2\left(h_{q_{2}}-h_{q_{1}}\right)-1}(e^{\frac{\pi x}{\tau_{0}}}+de^{\frac{2\pi x}{\tau_{0}}}) \nonumber \\
&- \frac{4a\gamma}{\pi}\ln\left(\frac{1}{d}\right)e^{2\pi i}(h_{q_{1}}-h_{w}-h_{v}) (h_{q_{2}}-h_{q_{1}}) (t-x)^{2\left(h_{q_{2}}-h_{q_{1}}\right)-1}\left(e^{\frac{\pi t}{\tau_{0}}}+de^{\frac{2\pi t}{\tau_{0}}}- (t \leftrightarrow -t)\right) \ , 
\end{align} 
In the limit of vanishing $a$, the above expression defines no interesting time-scale. However, when $a\not =0$, focussing only on the time-dependent part, the contribution due to inhomogeneity grows exponentially in time. However, since our expressions are valid only for $\tau_0 \gg t-x$, and $\tau_0 \gg x$, we cannot take the limit $t\gg \tau_0$. Thus, this exponential growth remains a transient behaviour.

For free fermions, the late time OTOC is dictated by a constant phase factor \cite{Kudler-Flam:2019kxq}. This can be argued simply due to factorization of four point OTOC into product of two point functions which differ from TOC by phase factors.  Hence, again we find no perturbative correction to OTOC due to inhomogeneous quench. Thus integrability of free fermions remains unaffected in this process in any time regime.

It is worth comparing the above examples with what happens in classical chaos. For classical integrable systems with $M$ degrees of freedom, the phase space is foliated by an $M$-dimensional tori. Here, the number of periodic directions, $M$, corresponds to the number of conserved charges. The celebrated KAM-theorem states that small, non-linear perturbations do not destroy this tori structure and therefor classical integrability is retained in the perturbative limit. For minimal models and free fermions, the OTOCs display a similar structure, for small perturbations. Moreover, the OTOCs lift the degeneracy between the dynamics of a Cardy-Calabrese state in a {\it chaotic} CFT ({\it e.g.}~large $c$ CFTs) and in a {\it super-integrable} CFT ({\it e.g.}~minimal models). This observation further highlights the importance of OTOCs in classifying dynamics, specially in the context of early-time chaos. We will now discuss an example that is somewhere in between a {\it chaotic} CFT and a {\it super-integrable} CFT.

\subsection{Orbifold CFT} \label{sec4.3}

Correlation function of twist operators and its growth has been extensively studied in \cite{Caputa:2017tju} to understand growth of Renyi entanglement entropy for subsystems in QFTs. In the cyclic orbifold CFT, these twist operators exist as natural candidates of primary operators. In \cite{Caputa:2017rkm}, OTOCs for twist operators in the orbifold CFTs has been studied as an attempt to characterize different CFTs in terms of chaotic properties alongside entanglement growth. In this section, we will use these results to study the change of OTOC due to inhomogeneous quench.

In concrete term, we consider free $c=2$ CFT on cyclic orbifold $(T^{2})^{n}/\mathbb{Z}_{n}$, where $T^{2} = S^{1}\times S^{1}$(or, two free bosons are compactified on the same radius $R$). Furthermore, let us take $W$ and $V$ as twist operators $\sigma_{n}$ with dimension $h_{n}=\bar{h}_{n} =\frac{1}{12}\left(n-\frac{1}{n}\right)$. The corresponding OTOC depends on the compactification parameter $\eta=R^{2}$. In the limit $t\gg x$, one can get:
\begin{align}\label{orbifold}
F &= \frac{1}{pp'}\ , \quad  \eta = \frac{p}{p'}\ , \quad  pp' \in 2\mathbb{Z} \ , \nonumber \\
&=0 \ , \quad  \eta=\frac{p}{p'}\  ,\quad  pp' \in 2\mathbb{Z}+1 \ , \nonumber \\
&= -\frac{\pi}{2\log\left(-\frac{\epsilon^{*}_{12}\epsilon_{34}}{16}e^{-\frac{\pi(t-x)}{2\tau_{0}}}\right)} \ , \quad \eta \neq \frac{p}{p'} \ . 
\end{align}
For rational $\eta$, the late time OTOC becomes constant (equivalent to rational CFTs) The only interesting late time behavior arises for irrational $\eta$ where the OTOC shows a power law decay as in (\ref{orbifold}). This result for irrational radius can be interpreted as a notion of `weak chaos',\footnote{Note that our use of  the term ``weak chaos" is different from \cite{Kukuljan:2017xag}.} since it does not show an exponential growth as in large $c$ CFTs or a constant behaviour like integrable rational CFTs or free theories. Furthermore, there is no interesting time-scale (associated to the {\it weak chaos}) other than $\tau_0$ that sets the universal dissipation scale in critical quench dynamics.

Let us now consider the effect of a small inhomogeneity. To see the correction, we first expand $F$ in (\ref{orbifold}) as:
\begin{align} \label{orbifoldF}
F \approx -\frac{\pi}{2\Sigma}\left(1+\frac{\pi}{2\Sigma\tau_{0}}(t-x)\right); \; {\Sigma\equiv \log\left(-\frac{\epsilon_{12}^{*}\epsilon_{34}}{16}\right)} \ , 
\end{align}
where the above expansion can be thought of as an expansion in the limit $\Sigma \to \infty$.\footnote{Recall that we can translate the auxiliary parameter $\Sigma$ to a physical one, by smearing the localized operators $V$ and $W$ over an Euclidean time-width. This width will serve the role of a regulator. This expansion will allow us to further take the $t\gg \tau_0$ limit. } Now, the VWS OTOC is obtained to be:
\begin{align}\label{scrambleorbifold}
&\frac{\langle V(\omega_{1},\bar{\omega}_{1}) W (\omega_{2},\bar{\omega}_{2}) V(\omega_{3},\bar{\omega}_{3})W (\omega_{4},\bar{\omega}_{4})\rangle \big|_{\rm VWS}}{\langle V(\omega_1,\bar{\omega}_1)V(\omega_3,\bar{\omega}_3)\rangle \langle W(\omega_2, \bar{\omega_2})W(\omega_4, \bar{\omega_4})\rangle\big|_{\rm VWS}}|_{\rm orbifold}\nonumber\\
&\sim -\frac{\pi}{2\Sigma}\left(1+\frac{\pi}{2\Sigma\tau_{0}}(t-x)\right) - \frac{a\pi}{2\Sigma^{2} \tau_{0}} \ln\left(\frac{1}{d}\right)\left[2e^{\frac{\pi x}{\tau_{0}}} - e^{\frac{\pi t}{\tau_{0}}} \right] \ .
\end{align}
When $a\not =0$, in the $t\gg x$ limit, there is an exponentially growing piece. This growth will subsequently take over the power-law growth in $t$, and subsequently define a scrambling time: $t_* \sim \tau_0 \log \left[ 1 /(-a \log d)\right] \gg \tau_0$, with $\lambda_{\rm L} = \pi /\tau_0$.  The above analyses can also be carried out without expanding in $\Sigma$. In this case, at early times, the denominator of (\ref{orbifoldF}) will be dominated by the constant term. Subsequently, (\ref{scrambleorbifold}) defines a scrambling time $t_*\gg \tau_0$. This example explicitly demonstrates that a {\it weak chaos} can be promoted to a standard early-time chaos defined by an OTOC, by inducing a small inhomogeneous perturbation.

\numberwithin{equation}{section}
\section{Conclusions}\label{sec5}

Our conclusions can be summarized into two main proposals. First, in critical quench dynamics, the universal effective thermal physics of the lower point functions\cite{Calabrese:2016xau} do generalize for a larger class of correlators, including the OTOCs for {\it chaotic} CFTs. However, OTOCs also distinguish a {\it chaotic} CFT from a {\it super-integrable} CFT. While we have used large-$c$ CFTs for the former and minimal models for the latter, as our examples, we expect these features to survive for a much wider range of systems. Our expectation is motivated from how CFT predictions qualitatively hold in a wider class of quench systems, see {\it e.g.}~\cite{Calabrese:2016xau} for some explicit examples.

From a Holographic perspective, the non-equilibrium dynamics of the critical homogeneous quench appears to be well-approximated by a suitable black hole geometry. Intuitively, this is similar to setting up a non-trivial boundary condition, of some field, at the AdS-boundary, and letting it evolve in time. If one now explores correlators at large time limit, due to Birkhoff's theorem, the dynamical geometry, away from the bulk fields, will be given by the Schwarzschild metric. This simple intuition transcends dimensions and hence it is suggestive of a broader applicability of the effective thermal description in dynamics.\footnote{Note, {\it e.g.}~in other steady-state configurations in Holography, similar effective thermal description emerges\cite{Das:2010yw}, \cite{Kundu:2019ull}.} Also, the basic assumptions\cite{Maldacena:2015waa} of cluster decomposition and analyticity are expected to hold for homogeneous quench as well. Taken together, our first proposal/conjecture is: For chaotic systems, under homogeneous quench, the corresponding Lyapunov exponent is upper bounded by the mass scale associated to the initial state. This conjecture is further consistent with numerical results in \cite{ddas}, which investigates a non-critical quench with Ising spins. Inhomogeneity provides some perturbative corrections to this physical picture. It will be an extremely interesting problem to prove a bound for these cases in general.

Our second proposal is related to how a ``weaker" integrable system, under generic deformation, becomes chaotic. While this is not surprising, explicit examples of such, in the realm of {\it early-time chaos} are hitherto unavailable. In this article, we explicitly demonstrated, using the integrable orbifold CFT, how an inhomogeneous quench leads to chaotic dynamics in this system. Moreover, the small inhomogeneity parameter defines a large scrambling time-scale that allows one to access this chaotic behaviour. While the example is specific, corresponding technical aspects are rather generic and are expected to hold for a wide class of CFTs with an order one central charge.

This CFT can furthermore be perturbed by a relevant operator, within a conformal perturbation theory. The corresponding OTOC will receive an additional contribution coming from a higher point function integrated over the entire complex plane. This correction, generally, will not cancel the inhomogeneity-induced exponential growth. Motivated by this, we propose/conjecture that non-trivial OTOC-dynamics for a generic {\it small-$N$} system can indeed be detected with an inhomogeneous quench set up. It would certainly be very interesting to explore this in a specific model in future, specially in the context of recent advances in experimental protocols of OTOC measurements, {\it e.g.}~in \cite{Sundar:2021lyv}.

From the perspective of quantum properties of black holes, it would be particularly interesting to understand how a Holographic calculation restores $v_{\rm B} =1$ for a general inhomogeneous quench. On the other hand, real-time dynamics of the dual CFT often encodes physics behind the horizon of the black hole, see {\it e.g.}~\cite{Hartman:2013qma} that captures universal features of time-evolution of entanglement entropy. Inhomogeneous quench is particularly interesting from this perspective, since a corresponding boundary data will likely affect the singularity structure inside the black hole. These will be very exciting future directions to further uncover.

{\bf Note Added:} We note that the upcoming paper \cite{ddas} has also independently obtained the Lyapunov exponent for homogeneous quench.

\section{Acknowledgements}

We thank Diptarka Das, Sumit Das, Shouvik Datta, Dileep Jatkar, Lata K.~Joshi, Sandipan Kundu, Krishnendu Sengupta for numerous illuminating discussions as well as comments on the draft. AK acknowledges support from the Department of Atomic Energy, Govt.~of India and IFCPAR/CEFIPRA project no.~6403. The work of SP and BR was supported by Junior Research Fellowship(JRF) from UGC. SD would like to acknowledge the support provided by the Max Planck Partner Group grant MAXPLA/PHY/2018577.


\appendix
\numberwithin{equation}{section}
\section{OTOC after at CWS in 2D CFT at large $c$}\label{A}

In this appendix, we will collect useful details for the large-$c$ CFT calculations, drawing heavily on the analyses of \cite{Roberts:2014ifa}. Towards that, we want to compute $F(\eta(\omega_{i},\bar{\omega_{i}}))$ at large $c$. Let us consider the case of $3$-point $b$-OTOC. The four points on the UHP after the analytic continuation are given by the following:
\begin{align}
z_{1}=-ie^{\alpha(t+i\epsilon_{1})}, \; z_{3}=-ie^{\alpha(t+i\epsilon_{2})}, \; z_{2}=-ie^{\alpha(x+i\epsilon_{0})}, \; \bar{z}_{2}=ie^{\alpha(x-i\epsilon_{0})}, \; \alpha= \frac{\pi}{2\tau_{0}} \ .
\end{align}
Here $\epsilon_{1} = \tau_{0}$ and $\epsilon_{2} = -\tau_{0}$. The choice of $\epsilon_{1,2}$ makes sure that when we take $t=0$, $z_{1,3}$ are still placed on the boundary of the UHP.\footnote{This we need for the standard analytic continuation from Euclidean to Lorentzian time. In doing so, primarily all the operators inside the Euclidean correlator are inserted at very small imaginary time i.e $\tau_{i} = i\epsilon_{i}$. Then we should continue the $\tau_{i}$s to their original Lorentzian values, i.e $\tau_{i} = t_{i}+i\epsilon_{i}$.} The corresponding cross ratio $\eta$ is:
\begin{align}\label{etacal}
\eta = \frac{2 e^{\alpha(x-t)}\left(e^{-i\alpha\epsilon_{2}}-e^{-i\alpha\epsilon_{1}}\right)}{(1-e^{\alpha(x-t+i\epsilon_{01})})(1+e^{\alpha(x-t-i(\epsilon_{2}+\epsilon_{0}))})} \ ,
\end{align}
where $\epsilon_{ij} \equiv \epsilon_{i}-\epsilon_{j}$.

Let us discuss analytic behaviour of $\eta$ in different time limits. For instance, at $t \rightarrow 0, t \gg x$ and $t=x$, we respectively obtain:
\begin{align}\label{z limit}
\eta_{t\rightarrow 0} = \frac{-2i\alpha\epsilon_{12}}{e^{\alpha x}-e^{-\alpha x}}\  , \quad \eta_{t\gg x} = 2ib\epsilon_{12}e^{-\alpha(t-x)} \  , \quad \eta_{t=x} = \frac{\epsilon_{12}}{\epsilon_{10}} = 1+\frac{\epsilon_{02}}{\epsilon_{10}} \ . 
\end{align}
Hence in the early and late time, the cross ratio $\eta \rightarrow 0$ from opposite direction in the complex $\eta$ plane. This behavior is completely independent of the time-ordering. However, at $t=x$, $\eta$ explicitly depends on the choice of ordering.

In this case, there exist two inequivalent time-orderings {\it i.e.}~$\epsilon_{0}>\epsilon_{1}>\epsilon_{2}$(time-ordered) and $\epsilon_{1}>\epsilon_{0}>\epsilon_{2}$(out-of-time-ordered). The ratio $\frac{\epsilon_{02}}{\epsilon_{10}}$ is $<0$ for time ordering and $>0$ for out of time ordering. Thus $\eta|_{t=x}>1$ for OTOC and $<1$ for TOC. This distinct behavior of $\eta$ plays a crucial role in determining the late time behaviour of the correlators, as we will review now.

In large $c$ limit, with $\frac{h_{w}}{c}$ is fixed and small while keeping $h_{v}$ large and fixed Virasoro vacuum identity block $\mathcal{F}(z)$ can be obtained analytically\cite{Fitzpatrick:2014vua}. Assuming further an identity block dominance, the full conformal four point function $f(z)$ can be approximated by the identity block contribution. The analytic expression of this block contains branch point at $\eta=1$ with the cut $[1,\infty)$. Following a contour around $\eta=1$ branch point and taking the small $\eta$ limit, one obtains\cite{Roberts:2014ifa}:\footnote{As we have shown earlier, the TOC cannot encircle the $z=1$ branch point and the contribution gives 1.}
\begin{align}\label{virasoro vac}
\mathcal{F}(\eta) \approx \left( \frac{1}{1-\frac{24i\pi h_{w}}{c\eta}}\right)^{2h_{v}} \ .
\end{align}
Now we substitute the large time behavior of $\eta$. Define $\Sigma_{ij} = i(e^{i\alpha\epsilon_{i}}-e^{i\alpha\epsilon_{j}})$ and $\Sigma^{*}_{ij} = -i(e^{-i\alpha\epsilon_{i}}-e^{-i\alpha\epsilon_{j}})$, the large time limit of $\eta$ becomes:
\begin{align}
\eta|_{t\gg x} = -2i\Sigma^{*}_{12}e^{-\alpha(t-x)} \ . 
\end{align}
Using this in (\ref{virasoro vac}), we get:
\begin{align}\label{Idblock}
F \approx \mathcal{F} &\approx \left(\frac{1}{1+\frac{12\pi h_{w}}{\Sigma^{*}_{12}}e^{ \alpha (t-t_{*}-x)}}\right)^{2h_{v}} \nonumber \\
&\approx 1- c' e^{\alpha(t-t_{*}-x)}, \; \text{for} \; t<t_{*} \quad c'=\frac{24\pi h_{w}h_{v}}{\Sigma^{*}_{12}} \ . 
\end{align}
Where the scrambling time $t_{*} = \frac{1}{\alpha}\log(c)$. Thus, we obtain the corresponding Lyapunov exponent:
\begin{align}
\lambda_{L} = \alpha = \frac{\pi}{2\tau_{0}} = \frac{2\pi}{\beta}  \ . 
\end{align}
Comparing with the usual maximal Lyapunov exponent at large $c$ {\it i.e.}~$\lambda_{\rm L} = \frac{2\pi}{\beta}$, we obtain an effective inverse-temperature is $\beta = 4\tau_{0}$. This is consistent with the effective temperature in homogeneous quench case as in the literature \cite{Sotiriadis:2008ila}.

For the bulk $4$-point function calculation, the computation proceeds in an identical manner, with the following operator insertion points on the CWS: $\omega_{1}= \omega_{3} = i\tau , $ $ \omega_{2}= \omega_{4} = x$.

\end{document}